\documentclass[a4paper]{spie}  

\usepackage[]{graphicx,amssymb}

\newcommand{\aap} {A\&A}
\newcommand{\araa} {Ann. Rev. A\&A}
\newcommand{\apjl} {ApJL}
\newcommand{\mnras} {MNRAS}
\newcommand{\apj} {ApJ}

\newcommand{\ao} {Applied Optics}
\newcommand{\phr} {Physics Reports}

\title{DEMON: a Proposal for a Satellite-Borne Experiment to study Dark Matter and Dark Energy}

\author{Alicia Berciano Alba\supit{a,b}, Pedro Borges da Silva\supit{c}, Hans Eichelberger\supit{d}, Francesca Giovacchini\supit{e,f}, Mareike Godolt\supit{g}, G\"unther Hasinger\supit{h}, Michael Lerchster\supit{i, h}, Vincent Lusset\supit{j}, Fabio Mattana\supit{k}, Yannick Mellier\supit{l}, Micha{\l} Micha{\l}owski\supit{m}, Carlos Monteserin-Sanchez\supit{n}, Fabio Noviello\supit{o}, \\Carina Persson\supit{p}, Andrea Santovincenzo\supit{q}, Peter Schneider\supit{r}, Ming Zhang\supit{s}, Linda \"Ostman\supit{t}
\skiplinehalf
\supit{a} Joint Institute for VLBI in Europe, postbus 2, 7990 AA Dwingeloo, The Netherlands;\\ 
\supit{b} Kapteyn Astronomical Institute, P.O.box 800, 9700AV Groningen, The Netherlands;\\
\supit{c} Funda\c c\~ao Navegar, Centro Multimeios de Espinho, Av. 24 $n^{\circ}$ 800 Espinho, Portugal;\\
\supit{d} Space Research Institute, Austrian Academy of Sciences, 8042 Graz, Austria; \\
\supit{e} Department of Physics, Bologna University, via Irnerio 46, I-40126 Bologna, Italy; \\
\supit{f} Istituto Nazionale di Fisica Nucleare, Viale B. Pichat 6/2, I-40127 Bologna, Italy; \\ 
\supit{g} Astrophysical Institut Potsdam, An der Sternwarte 16, 14482 Potsdam, Germany; \\
\supit{h} Max Planck Institute for Extraterrestrial Physics, Giessenbachstr., D-85748 Garching, Germany; \\
\supit{i} University Observatory Munich, Faculty for Physics at the Ludwig Maximillians University, Scheinerstr. 1, 81679 Munich, Germany;\\ 
\supit{j} Service de Physique des Particules, Commissariat \'a l'Energie Atomique - Saclay, bat 141 F-91191 Gif-sur-Yvette cedex, France;\\
\supit{k} Istituto Nazionale di Astrofisica, Istituto di Astrofisica Spaziale e Fisica Cosmica di Milano, Via Bassini 15, 20133 Milano, Italy; \\
\supit{l} Institut d'Astrophysique de Paris, 98 bis Boulevard Arago, F-75014 Paris, France; \\
\supit{m} DARK, University of Copenhagen, Juliane Maries Vej 30, 2100 Copenhagen, Denmark;\\
\supit{n} Instituto de Fisica de Cantabria, Avd. Los Castros s/n, E-39005 Santander, Spain; \\
\supit{o} Dept. of Experimental Physics, NUI Maynooth, Maynooth, Co. Kildare, Ireland; \\
\supit{p} Onsala Space Observatory, Chalmers University of Technology, S-439 92  Onsala, Sweden; \\
\supit{q} ESA/ESTEC TEC-SYE, P.O. Box 299, NL-2200 AG Noordwijk ZH, The Netherlands; \\
\supit{r} Argelander-Institut f\"ur Astronomie, Auf dem H\"ugel 71, D-53121 Bonn, Germany; \\
\supit{s} University of Manchester, Jodrell Bank Observatory, Macclesfield SK11 9DL, UK; \\
\supit{t} Department of Physics, Stockholm University, SE 106 91, Stockholm, Sweden;
}

\authorinfo{Further author information: (Send correspondence to A.B.A.)\\
A.B.A.: E-mail: berciano@astro.rug.nl, Telephone: +31 50 363 3505 \\
}

\begin{document} 
  \maketitle 

\begin{abstract}

We outline a novel satellite mission concept, DEMON, aimed at
advancing our comprehension of both dark matter and dark energy,
taking full advantage of two complementary methods: weak lensing and
the statistics of galaxy clusters. We intend to carry out a 5000 $\rm
deg^{2}$ combined IR, optical and X-ray survey with galaxies up to a
redshift of z $\sim$ 2 in order to determine the shear correlation
function. We will also find $\sim 100000$ galaxy clusters, making it
the largest survey of this type to date.  The DEMON spacecraft will
comprise one IR/optical and eight X-ray telescopes, coupled to
multiple cameras operating at different frequency bands. To a great
extent, the technology employed has already been partially tested on
ongoing missions, therefore ensuring improved reliability.
\end{abstract}



\section{Introduction}
\label{sec:introduction}

The open question concerning the nature of dark matter and dark energy
is certainly one of the most compelling issues of modern astrophysics,
that demands new and better observations in order to be
answered. Until the physical properties of dark matter can be
determined experimentally (something that cannot be done in the
laboratory up to now), trying to improve our knowledge about these properties
implies improving the current measurement of the cosmological
parameters. In particular, in the case of dark energy, the first
crucial step is to try to distinguish a cosmological constant model
from time dependent quintessence models.

The current best constraints on the cosmological parameters come from
the combined results of three different techniques with complementary
degeneracies: SNIa experiments, CMB measurements, and the study of
galaxy clusters.  In addition to these well known methods, weak
lensing is emerging as one of the most powerful and robust probes of
the large scale structure of the universe and the nature of
dark matter and dark energy\cite{mellier, bartelmann, refrieger}.

Several existing and proposed wide-field weak lensing surveys,
ground-based\footnote{VISTA (Visible \& Infrared Survey Telescope for
Astronomy), Pan-STARRS (Panoramic Survey Telescope And Rapid Response
System), LSST (Large Synoptic Survey Telescope), CFHTLS
(Canada-France-Hawaii Telescope Legacy Survey), KIDS (Kilo-Degree
Survey) and DES (Dark Energy Survey)} and space borne\footnote{JDEM
(Joint Dark Energy Mission) from NASA-DOE and DUNE (Dark UNiverse
Explorer) from CNES, also proposed for ESA's Cosmic Vision} are
currently being developed. In order to contribute to the design of
possible space missions we present DEMON (Dark Energy and Matter
Observational Nexus), a novel concept satellite conceived to study the
properties of dark matter and dark energy combining two complementary
techniques: weak lensing and clusters statistics.


\section{Overview: Weak Lensing and Cluster Statistics}
\label{sec:goal}

In order to address the above science questions, DEMON will carry out
a combined optical/IR and X-ray wide-field survey to provide the
required data for weak lensing and cluster statistic studies. Sections
\ref{sec:wl} and \ref{sec:cs} explain the basis of both techniques and
the usefulness of combining them to study the problem of dark matter
and dark energy.

\subsection{Weak lensing}
\label{sec:wl}

The weak gravitational lensing effect is the deflection of light by
large scale structures along the line-of-sight. This produces a
coherent distortion of the shape of observed background galaxies
(cosmic shear), which contains information about the three dimensional
mass distribution of the universe and also about the structure growth with
time. The cosmic shear maps obtained from weak lensing surveys can be
used to reconstruct the dark matter power spectrum, prove the effect
of dark energy in the geometry of the universe (geometrical tests),
and measure the cosmological parameters analyzing the 3D shear power
spectrum in redshift slices (tomography).  These maps can be obtained
through the statistical properties of the observed ellipticity of
galaxies, which requires a large statistical sample of galaxies and very
good image quality to measure the galaxy ellipticities reliably.

Since the cosmic shear is only due to gravity, weak lensing does not
suffer from galaxy bias uncertainty like the baryon oscillation
technique, and is not affected by evolution and/or environmental
effects as may be the case for SNIa.  In addition, weak lensing offers
a unique method to probe the matter distribution in the transition
domain from quasi-linear to non-linear scales at medium redshifts
which cannot be addressed by any of the other techniques. Studies of
cosmic shear already have led to significant constraints on
cosmological parameters, most noticeably on the normalization of the
power spectrum\cite{jarvis}.

\subsection{Cluster statistics}
\label{sec:cs}

Clusters of galaxies are the largest and most recently collapsed
objects in the universe. While their formation and evolution are
driven by gravity, their large scale distribution and growth rate are
also affected by dark energy. Two methods involving clusters are
especially promising in cosmology: determining the redshift evolution
of the matter power spectrum, and measuring the redshift evolution of
the cluster mass function (the abundance of clusters as a function of
mass) \cite{haiman}. Comparing these statistical quantities with those
predicted by structure formation theory allow us to measure the
cosmological parameters and the equation of state of dark energy
\cite{voit}.

Cluster statistics require mass determination for each cluster, which
can be inferred from observables such as the intracluster gas
temperature in the X-ray spectrum, and the bolometric X-ray luminosity
\cite{reiprich}.
Some X-ray surveys of galaxy clusters have measured the power spectrum
and the cluster mass function \cite{rosati}, but not its evolution with
redshift. Missions like XMM-Newton with a random survey precludes any
studies based on spatial correlation (such as the matter power
spectrum). These effects are even more important for the Chandra X-ray
Mission and the planned XEUS (X-ray Evolving Universe Spectroscopy
Mission) and Constellation-X mission, with their smaller fields of
view.

In order to reach estimates with errors of a few percent, observations
of ten thousand galaxy clusters in contiguous fields and with well
defined completeness are needed\cite{haiman}. This goal can be
efficiently accomplished by a wide-field X-ray survey. In addition,
the mass$-$luminosity/temperature relations needed for cluster
statistics studies can be considerably improved with additional mass
measurements \cite{majumdar} provided by alternative methods like weak
lensing \cite{pedersen}. Since weak lensing and cluster statistics
also provide complementary measurements of the cosmological
parameters, it is very suitable to combine them in the same mission.


\section{Science case}
\label{sec:requirements}

The main goal of the DEMON satellite is to provide optimum quality
data for weak lensing studies, the scientific requirements are
discussed further in Sect.~\ref{sec:weaklensreq}. On the other hand,
since cluster statistic studies are an interesting additional feature
for the mission, we intend to take advantage of the X-ray telescopes
developed for the eROSITA mission to carry out a simultaneous X-ray
survey. The characteristics of such an X-ray survey are explained in
Sect.~\ref{sec:xrayreq}.

\subsection{Weak lensing scientific requirements}
\label{sec:weaklensreq}

Since the weak lensing technique is based on measurements of galaxy
shapes, high image quality is a mandatory requirement in order to
achieve optimal results. That makes a space-based wide-field survey
the natural option to overcome the limitations produced by the
degradation of the images in ground based surveys (smearing due to
atmospheric seeing, PSF instabilities, weather instabilities and
telescope instabilities due to flexures and wind-shake). A space
telescope provides a diffraction limited and stable point spread
function (PSF) minimizing the systematics of the images, and allowing
the detection of more faint resolved galaxies.

In order to provide a high sample of galaxies to minimize statistical
errors, we need a large sky coverage and a high number density of
resolved galaxies for shape measurements\cite{tyson}. With a limiting
magnitude of $\rm I_{AB} \sim 27$, a galaxy density of $\sim 60-100
~\rm arcmin^{-2}$ is
expected\cite{aldering,heymans,schneider,schrabback,tyson}. This
provides a galaxy sample with $\langle z \rangle \sim 1.3-1.4
$\cite{rhodes,heymans}, suitable to probe the range\cite{ferguson}
$z=0-2$ in which the effect of dark energy is more
prominent\cite{kirshner}. According to Tyson et al.\cite{tyson},
ellipticity values can be retrieved for galaxies larger than
1.2$\times$PSF. Since galaxies with $\rm I_{AB} \sim 27$ have mean
sizes ($2\times$half-light radius) of $\sim0.2''$ \cite{schneider}, we
need a telescope with a full width half maximum (FWHM) of the PSF $\le
0.17''$. In addition, accurate measurements of galaxy shapes require a
FWHM sampling of at least $2\times2$ pixels for a proper PSF
deconvolution.

To assess the above scientific requirements, we studied a baseline
model consisting of a 2 meter telescope similar to the one suggested for
SNAP\cite{snap_telescope} (SuperNova/Acceleration Probe). We intend to
cover 5000 $\rm deg^{2}$ with an exposure time of 4 ks per
filter. The instrumentation requirements are described in
Sect.~\ref{sec:opt_inst}.


Weak lensing tomography also requires the determination of the
photometric redshifts of the galaxy sample (too big for spectroscopic
redshift measurements) which can be obtained with a targeted accuracy
of the order of $0.03(1+z)$ using 8 filters \cite{bolzonella}. Since
it is not practical to include them all in a single spacecraft, a
ground-based optical wide-field imaging survey done in parallel will
be needed to achieve the mentioned accuracy. The minimum requirement
for the ground-based survey is a 4 meter diameter optical telescope with a
3 deg$^2$ field of view (FOV) operating in the U, B, V, and R
bands. Projects of this kind, such as VISTA, KIDS, Pan-STARRS...
are currently being developed.  The IR filters should be included in
the spacecraft, since the sensitivity of ground-based IR observations
is severely limited by atmospheric and thermal effects.

\subsection{X-ray survey characteristics}
\label{sec:xrayreq}

The kind of X-ray survey that we can perform with DEMON is determined
by: (1) the sky coverage (5000 $\rm deg^{2}$) and exposure time
(16 ks per pointing) imposed by the weak lensing mission
requirements, and (2) the technical specifications of the X-ray
telescopes developed for the eROSITA mission (see
Sect.~\ref{sec:xray_inst} for details).

The eROSITA telescopes are optimized for cluster statistics
studies. In particular, they provide an angular resolution of $<15''$,
suitable to distinguish between galaxy clusters and point-like X-ray
sources\cite{mullis}.  With an energy resolution of 130 eV at 6 keV,
they also provide an X-ray spectrum with enough resolution to measure
the mean cluster gas temperature.

An X-ray instrument composed by 8 of such a telescopes will have a
total FOV of 3.55 $\rm deg^{2}$. Combined with an exposure time of
$\sim 64$ ks per pointing, such an instrument will reach a minimum
flux of $8\times10^{-15}$ erg cm$^{-2}$ s$^{-1}$ in the energy range
$0.5-2$ keV. Under this conditions, the DEMON X-ray survey is expected
to detect $\sim$ 20 clusters per~$\rm deg^{2}$ and identify, in the
observed region of the sky, all the clusters with a mass $\gtrsim$ $2
\times 10^{14}$ $\rm M_\odot$ up to $z \sim 1.5$. Therefore, DEMON
will produce a catalog of $\sim 100000$ clusters with a mean redshift
of 0.75. The flux-selection will avoid the contamination from clusters
highly affected by non-gravitational heating, mainly non massive and
non relaxed\cite{rosati}.


The study of the evolution of the matter power spectrum and cluster
mass function requires an error $\sim 0.02$ in the cluster redshift
measurements\cite{haiman}. This information will be provided by the
DEMON (combined with the required ground-based survey) with a median
accuracy of 0.07 for the redshift of each galaxy. Therefore, redshift
measurements of $\sim 12$ galaxies per cluster will be enough to
achieve the required precision.


Another interesting feature of the DEMON compared to other X-ray
  surveys is that, due to the flux limit and survey area that is going
  to cover, it will explore a measurement domain that was not done
  before.

\section{Instrumentation}
\label{sec:intrumentation}

The baseline optical concept will comprise a 2 m diameter
optical/IR telescope encircled by eight 0.358 m diameter X-ray
telescopes. Below, the optical/IR and X-ray telescopes and detectors
are described. The specifications of both instruments are summarized
in Tables \ref{tab:opt_telescope} and \ref{tab:xrayparam}.

\subsection{Optical/IR telescope and detectors}
\label{sec:opt_inst}

\begin{figure}
   \begin{center}
   \begin{tabular}{c}
   \includegraphics[height=10cm, angle=-90]{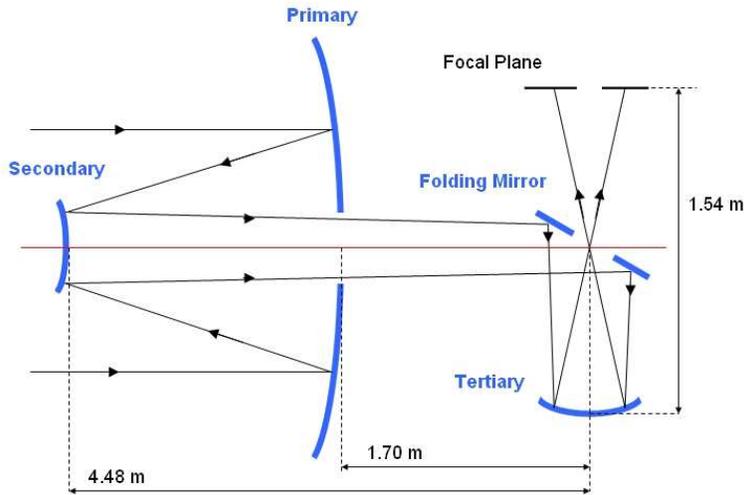}
   \end{tabular}
   \end{center}
   \caption {The optical mirror system (not to scale), without
   magnifying lens system. }
   \label{fig:mirror} 
\end{figure}

The main DEMON optical/IR instrument is a three-mirror anastigmatic
Korsch telescope\cite{korsch} (see Figure~\ref{fig:mirror}). This
configuration produces an achromatic image limited by diffraction over
a planar field of more than 1 $\rm deg^{2}$, and has been used extensively
in space. Honeycomb sandwich mirror technology is chosen to minimize
the mass of the mirror.

DEMON will be equipped with 4 filters: one optical broad-band filter
 to measure the galaxy shapes (R+I+z, centered at 0.8 $\mu$m with 0.2
 $\mu$m width), and 3 IR filters to determine the photometric
 redshifts (J, H and K).

In the broad band filter, the size of the telescope FWHM PSF is
$0.08''$, which requires an angular resolution of $0.04''$/pixel (to
be properly sampled) and a pixel size of 5 $\mu \rm m$. Since this
pixel size is much smaller than the current technology available, the
DEMON pointing strategy will take advantage of
micro-dithering\cite{schuler}-- the integration time is divided into
two sub-pointings separated by half a pixel. This allows us to use a
pixel $\sqrt{2}$ times larger. Combined with a magnifying lens system
(to enlarge the image by a factor of 1.5), the PSF can then be
properly sampled using 10 $\mu \rm m$ pixels. Since there is no need
for this high resolution for the infrared filters, we can use IR
detectors with the standard pixel size of 18 $\mu \rm m$
($0.085''$/pixel).

\begin{figure}
   \begin{center}
   \begin{tabular}{c}
   \includegraphics[height=7cm]{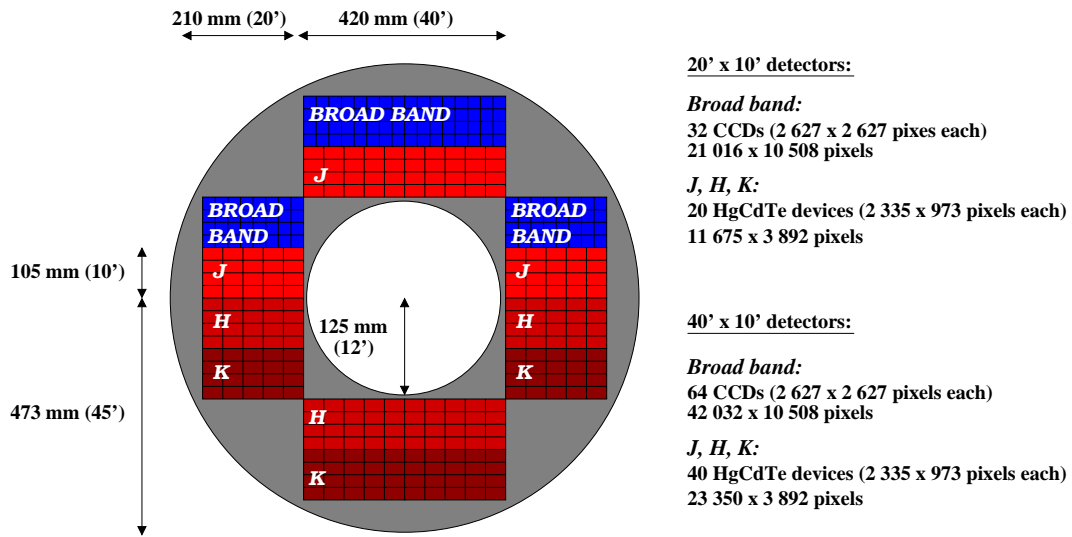}
   \end{tabular}
   \end{center}
   \caption {Optical/IR focal plane and detectors.}
   \label{fig:detector} 
\end{figure}

The DEMON focal plane is a ring with a partially vignetted area in the
center, where the detectors are situated in four 20' $\times$ 40'
rectangular areas (see Figure~\ref{fig:detector}). The four filters are
settled in one direction to cover the same FOV, which
requires an orthogonal scanning strategy. In total DEMON will have 128
CCDs (2627$\times$2627 pixels each) for the broad-band filter, and 80
IR pixel arrays (2335$\times$973 pixels each) for each of the IR filters.

The SNAP collaboration have successfully developed a new type of large
format back-illuminated p-channel CCDs with 10.5 $\mu \rm m$ pixels,
specially suitable for space\cite{gigacam}. Therefore, the development
of 10 $\mu$m pixels will not be a technical problem. For the IR
detectors we can use HgCdTe devices from Rockwell with a 2.5 $\mu$m
wavelength cut-off.

\begin{table}
\begin{center}
\caption{Specifications of the optical/IR system.}
\label{tab:opt_telescope}
\begin{tabular} {p{4.5cm} p{8cm}}
\hline
\hline
Parameter  & Value \\
\hline
Telescope                 & 3-mirror system  \\
Focal length              & 24 m  \\
Effective focal length    & 36.12 m  \\
Primary mirror diameter   & $2000$ mm   \\
Primary mirror thickness  & 400 mm \\
Secondary mirror diameter & 466.7 mm   \\
Tertiary mirror diameter  & $1066.7$ mm   \\
Fold mirror diameter      & 489.3 mm   \\
Baffle length             & 0.7 m \\
FOV                       & $0.9 \rm {deg}^{2}$ (0.23 $\rm deg^{2}$ each filter)\\
Optical angular resolution & 0.04''/pixel (at 0.8 $\mu$m) \\
IR angular resolution      & 0.10''/pixel \\
Image stability           & 0.03''  \\
Filters                   & Broad-band ($0.6-1 \mu \rm m$), J, H, \& K  ($600-2500$ nm) \\
Optical detector          & 2.6k $\times$ 2.6k CCDs of 10 $\mu$m pixels \\
IR detector              & 2.3k $\times$ 973 HgCdTe devices of 18 $\mu$m pixels \\
\hline
\end{tabular}
\end{center}
\end{table}

\subsection{X-ray telescope and detectors}
\label{sec:xray_inst}

The DEMON X-ray instrument consists of eight X-ray telescopes with a
design similar to that developed for the eROSITA mission. Telescope
specifications are explained in detail in the eROSITA Mission
Definition Document (www.mpe.mpg.de/erosita/MDD-6.pdf).

Basically, each X-ray telescope is composed of 54 nested Wolter-I
mirror shells (paraboloid+hyperboloid optics), with 27 outer shells to
enhance the effective area at low energies. The optical design of the
X-ray mirror system has hexagonal geometry, optimized to achieve
maximum sensitivity between 0.5 and 10 keV. The mirrors are fabricated
using a nickel-galvano plating, coated with gold to enhance their
reflectivity. Each mirror module has its own CCD-detector, mounted in
its own housing and equipped with its own electronics. The CCD size of
$19.2\times19.2$ mm$^2$ corresponds to a FOV of
$41.2'\times41.2'$. The X-ray detectors are based on the pn-CCD
principle, although they have a smaller pixel size ($75\times75$
$\mu$m$^2$) and shorter readout time than regular pn-CCDs. The
semiconductor laboratory of MPE are already fabricating these kind of
detectors.

\begin{table}
\begin{center}
\caption{Specifications of the X-ray system.}
\label{tab:xrayparam}
\begin{tabular}{l l}
\hline
\hline
Parameter & Value \\
\hline 

Focal length & 1600 mm \\
Diameter of one mirror system  & 358 mm \\
Baffle length & 600 mm \\
Telescope size (diameter/length) & 0.4 m /2.6 m \\
FOV                & 0.473 $\rm deg^{2}$ \\
Angular resolution & $< 15''$ (at 1 keV) \\
CCD size & $19.2\times19.2$ mm$^2$ \\
Pixel size & 75 $\times$ 75 $\rm \mu$m$^2$ \\
Readout time & 50 ms \\
Energy range & $0.5-10$ keV \\
Energy resolution & 130 eV at 6 keV \\
Effective collecting area & 2656 cm$^{2}$ \\
Total Grasp & 753.6 cm$^{2}$deg$^{2}$ (at $1$ keV) \\
\hline
\end{tabular}
\end{center}
\end{table}


\section{Spacecraft orbit and scanning strategy}
\label{sec:scan}

In order to achieve the scientific objectives of the mission and to
meet the spacecraft's operational requirements, we have chosen a
circular low-earth orbit (LEO) with the parameters given in
Table~\ref{tab:orbit}.  This particular orbit was chosen to avoid the
Van Allen radiation belts and the South Atlantic Anomaly, both of
which would prejudice X-ray observations. The selected LEO will also
facilitate ground control and data download procedures. Because of
atmospheric drag, periodical corrective maneuvers (carried out by
thrusters) will be needed to reposition the spacecraft into its
desired orbit. The propellant mass required for these maneuvers has
been computed at 500~kg (for a 5 year mission). In total, the
spacecraft's dry mass is 2000 kg (20\% marging included), which fits
well within the transport capabilities of 5500 kg of a Soyuz rocket
launched from Kourou into the desired LEO orbit (see
Table~\ref{tab:mass_power} for details).

\begin{table}
\begin{center}
\caption{Orbital parameters.}
\label{tab:orbit}
\begin{tabular} {l l}
\hline
\hline
Parameter & Value \\
\hline
$\Theta_{\rm inclination}$	& 5 deg \\
Altitude	& 600 km \\
Velocity	& 7.6 km s$^{-1}$ \\
Period	& 1.61 h \\
$t_{\rm eclipse}$	& 28 min \\
\hline
\end{tabular}
\end{center}
\end{table}

The telescopes will be allowed to shift within a cone aligned with the
Earth's rotational axis with a maximum semi-aperture angle $\theta_{\rm
sa}=21$ deg. This will avoid bringing the Earth, the Sun and the plane
of the Milky Way into the instrumental FOV. The exact pointing range
will also depend on the specific position of the Earth along its
orbit. The selected value of $\theta_{\rm sa}$ allows, in principle,
the coverage of about 15000 deg$^2$ of the sky around the Earth's
north and south poles.
The sky will be sampled in concentric circles while the center of each
FOV will be shifted by $10''$ every two orbits in accordance with
detector array technical requirements.

With an exposure time of 16000 s for each pointing, this
configuration is expected to reach the scientific demands for a 5000
$\rm deg^{2}$ survey in a 4.5 years mission (including 68\%
efficiency).

%
%
%
%


\begin{table}
\begin{center}
\caption{Estimated masses and power requirements of the spacecraft components.}
\label{tab:mass_power}
\begin{tabular}{l l l }
\hline
\hline
Subsystem                          & Mass    & Power\\
\hline
Propellant                                       & 500 kg  &\\
Spacecraft structure \& Harness                  & 280 kg  &                \\
Telescope structural support                     & 160 kg  & \\
Optical mirror system                            & 120 kg  &  \\
Optical/IR detectors                             & 200 kg  &  480 W \\
X-ray telescope \& detectors                     & 700 kg  &  110 W  \\
Power system                                     & 250 kg  &  130 W \\
Thermal control \& Multi-layer insulation        & 100 kg  &  900 W  \\
Attitude control                                 & 90  kg  &  250 W \\       
Electronics \& Telemetry                         & 100 kg  &  110 W \\
\hline
{\bf Total}                        & {\bf 2500 kg} & {\bf 1980 W}  \\
\hline
\end{tabular}
\end{center}
\end{table}




\section{Spacecraft Engineering}

\subsection{Spacecraft design}

The DEMON spacecraft is divided in two main parts: the main body and
the service module.

The main body is a 6 m high cylinder with 2.6 m of diameter. It
contains the optical/IR and X-ray telescopes and detectors, the solar
panels, and a radiator to cool the X-ray focal planes. To avoid any
direct stray light into the mirror system, the optical/IR telescope
will be protected by a cylindrical deployable baffle with a total
length of 110 cm (70 cm outsize the main tube of the telescope). Each
X-ray telescope has also its own cylindrical baffle with a total
length of 60 cm.

The service module is an octagonal box which is 1.1 m high and 3.75 m
in diameter. The lateral sides include the phased array antenna
system. The bottom side has two antennas, the Soyuz connection ring
and radiators. At the sides of the service module there are two solar
sensors, two star trackers and 12 thrusters. Inside the service module
there is the thermal control system, the attitude control system, the
power conditioning devices, the electronic devices, the telemetry and
telecommand devices and the batteries.

\begin{figure}
\centering
	\includegraphics[width=12cm]{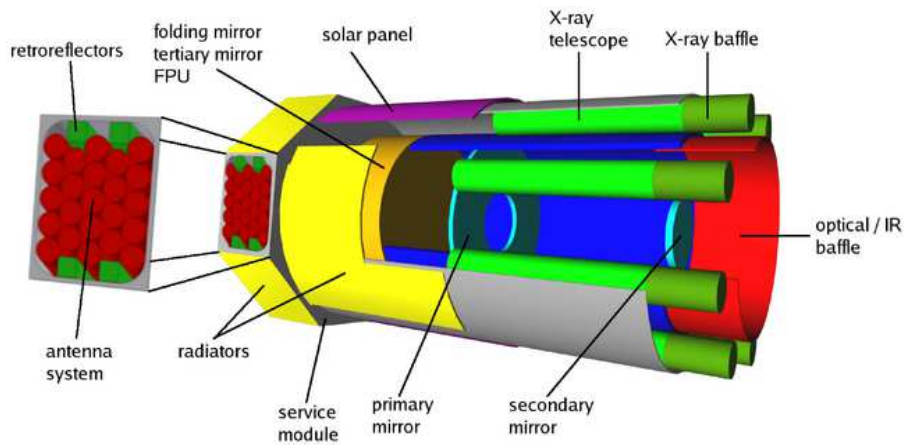}
\caption{The design of the spacecraft.}
\label{fig:design}
\end{figure}

An artist impression of the satellite is shown in Fig. \ref{fig:design}.

\subsection{Power system}

During most of the orbit, the power of the spacecraft will be supplied
by a solar array built from ultra triple junction solar cells of
GaInP$_2$, GaAs, and Ge. The solar array will also charge batteries
that will be used during the eclipses and in particular circumstances
when greater amounts of power are needed. Lithium-ion batteries will
be used since these have a large interval of allowed operational
temperatures ($248 \rm ~K - 268 \rm ~K$) facilitating thermal control.

The power needed for the different subsystems of the spacecraft are
listed in Table~\ref{tab:mass_power}. It amounts to a total required
power of 1980 W. The necessary size of the solar array is therefore
14.7 m$^2$ taking into account the losses of energy during
conversions, the cell efficiency and deterioration of components with
time.
\begin{table}
\begin{center}
\caption{Estimated costs for a five years mission.}
\label{table:costs}
\begin{tabular}{l l}
\hline
\hline
Part & Cost (in M euro)\\
\hline
Main telescope	    & 50 \\
Optical/IR camera   & 80 	\\ 
X-ray instrument    & 16 	\\
Platform	           & 200  \\
Launcher	           & 50   \\  
Ground support	    & 100 ~(20 per yr) \\  
Ground observations & 12.5   ~(2.5  per yr) \\
\hline
{\bf Total}   &{\bf 508.5 }  \\
\hline
\end{tabular}
\end{center}
\end{table}

\subsection{Thermal control}

In order to minimize the radiative exchange and ensure the
insulation, the satellite is wrapped with layers of aluminized mylar
and covered with a special white paint.

Due to instrumentation requirements, the service module and payload
operate at different temperatures.  For this reason they must be kept
thermally insulated from each other. The service module is maintained
at a temperature between 278 K and 303 K using a passive cooling
system consisting of two radiators. The radiators are placed on the
cold side of the service module to carry off the heat produced inside
by the detector's electronic boards, attitude control, propellants,
power and communication system and electronic internal dissipation.
The batteries are thermally isolated and equipped with heaters
controlled by thermostats. The two star trackers are cooled with
Peltier coolers.

\subsubsection{Thermal control of the IR detector}

Active cooling is required for the IR detector since the focal plane
has to be kept at 150 K. For this purpose there are 16 thermoelectric
pumps and 32 spares. Each pump has an absorption capacity of 1.22 W
and a required power supply of 43.5 W. A radiator is placed on
the cold side of the spacecraft to radiate this power.

\subsubsection{Thermal control of the X-ray detector}

The focal plane of each X-ray detector has to be maintained at an
operating temperature of 213 K in order to achieve optimal
functionality and energy resolution. Since the adopted attitude
strategy maintains one side of the spacecraft hidden from the sun and
the earth (except for 1.5 minutes per orbit), passive cooling is a
possible solution. The total heat flux coming from the albedo and IR
radiation of the earth at that cold side was computed with the ESARAD
software for different positions of the spacecraft along the
orbit. Based on this result, we plan to use radiators and heat-pipes to
cool the X-ray detectors. To avoid overheating the CCD focal plane
during the hot cases along the orbit, active elements will support the
cooling.  Since the mirror system operate at ambient temperature,
heaters with thermostats are mounted on the mirror modules in order
to maintain temperature stability (e.g. $293\pm2$~K).

\subsection{Attitude control}

The thermal control requires two 180$^\circ$ slews of the spacecraft
every orbit: a direct slew, and a retrograde slew to return the
spacecraft to its initial position (similar to the strategy adopted
by eROSITA). Due to the short period of the orbit, the slew rate
should be of the order 1$^\circ$/sec (a quarter spin must be done in
less than two minutes).

To accomplish the survey strategy, the spacecraft should be able to
make slews in all three directions. Therefore, 4 momentum wheels (one
for each axis) are used for the actuators. The spacecraft includes a
spare momentum wheel in case one of the others breaks
down. Twelve hydrazine thrusters are used for momentum dumping,
orbital corrections and to assist in more demanding maneuvers, such as
the 180$^\circ$ slew.

To guarantee a high spacecraft stability, we have three different
types of attitude control movements: maneuver, pointing and stability
mode. For this purpose three types of sensors are used: two sun
sensors, two star trackers, and two space inertia reference units. The
maneuver mode consist of retrieving data from the two sun sensors and
performing a fast slew. In the pointing mode, the star trackers permit
more precise maneuvers. Finally, the stability mode is required for long
time exposures when maximum stability is necessary. In this case, the
space inertia reference unit (that uses the data from six gyroscopes)
gives a precision of one arcsecond per hour. The use of a kilo-Hertz
satellite laser ranging system as a backup has also been envisaged.



\subsection{Communication}

The DEMON communication, telemetry tracking and command (TT\&C)
segment will be based on reliable radio frequency (RF) technology in
the Ka-band, since this is the only band which will be able to offer
sufficient telemetry capacity. In a few years, the Ka-band technology
will be fully space qualified and tested on several space-borne
missions. The specifications of the memory and telecommunication
subsystems are shown in Table~\ref{demon_telecom}.

\subsubsection{Space segment}

The baseline concept for the science and telemetry link are compact
and efficient Ka-band solid-state power amplifiers (SSPAs), together
with a ``semi-active'' conformal phased array antenna, Butler-like
matrices, and MMIC phase shifters \cite{vourch,caille,polegre}.
The antenna has no moving part, 
as demanded for extremely fine pointing accuracy.
The power transferred from the antenna system to the spacecraft is
kept constant in order to minimize negative effects from a temperature
change into the optical payload.

Science 
and engineering data are buffered in the solid-state mass memory
(SSMM) on board, and loss-less compressed with a consultative committee
for space data systems (CCSDS) compliant solution.  This data is
subsequently down-linked via the telemetry system and antenna to the
ground station.

\subsubsection{Ground segment}

The orbit of DEMON enables the usage of ground stations with low cost,
commercial Ka-band front-ends in Kourou/French Guiana and
Malindi/Kenya.
Beside the broadband telemetry link from the spacecraft to the ground
stations, data relay system satellite services will be used with
inter-orbit links (IOL) in Ka-band
for telecommand and real-time engineering telemetry.

DEMON will also use several link management techniques for optimization,
such as site-time diversity with selective re-transmission, short term
forecast driven weather adaption, and error control coded operations.

\begin{table}
\begin{center}
\caption{Parameters of the memory and telecommunications subsystem.}
\begin{tabular}{ll}
\hline
\hline
Parameters, Hardware                                     & Values, Annotations                           \\
\hline
Frequency of the Ka-band TT\ensuremath{\&}C                               & 25.5-27 GHz (alternatively 37-38 GHz)         \\
Bandwidth efficient modulations \cite{CCSDS_413_0_G_1}   & OQPSK/PM, OQPSK I/Q, SOQPSK, FQPSK-B, etc. \\
Forward error correction codes                           & Turbo Code, Reed-Solomon; FPGA implementation                      \\
SSPAs                                                    & max. 1 W\ensuremath{_{RF}} each, in all 60 W\ensuremath{_{DC}} for power system  \\
Telemetry data rate                                      & 500 Msps up to 1.2 Gsps (fair weather)        \\
SSMM with lossless data compression                      & min. 500 GBit at end-of-life; on-board SAN                 \\
CCDs uncompressed raw data rate                          & max. continuous 160 Mbps                                      \\
Data rate for SSMM / Compression factor                  & 60 Mbps / $\sim$ 3               \\
Ground stations 
& Kourou
and Malindi,
each with 500 s of visibility per orbit          \\
\hline
\end{tabular} 
\label{demon_telecom}
\end{center}
\end{table}


\section{Conclusions}
\label{sec:conclusions}

This paper presents a new concept mission for a satellite aimed at
provide better constraints on the cosmological parameters and, in
particular, studying the evolution of the dark energy equation of
state with time.

The innovate concept of the DEMON spacecraft is to carry out a
combined optical/IR and X-ray survey of 5000 $\rm deg^{2}$ to perform
weak lensing and cluster statistics studies. The satellite is meant to
be complemented by one of the ground-based wide-field surveys
currently under development, to provide the best accuracy in the
photometric redshift measurements required.

For the optical/IR part of the survey, we investigated telescope
design similar to the one proposed by SNAP.  To sample the FWHM PSF
with $2\times2$ pixels (something crucial for weak lensing studies),
the proposed telescope requires a pixel size of 5 $\mu m$. This
technology is currently under development and is expected to be
available within 5 years (J. W. Beletic, director of sensor systems of
Rockwell Scientific, private communication).  An alternative possibility
consist on use micro-dithering in the pointing strategy and include
an additional magnification of 1.5 in the optical system. This
configuration allows in principle to sample the FWHM PSF with
$2\times2$ pixels using CCDs of 10 $\mu$m. However, the derived
consequences in the technical telescope design and image quality
should be investigated in detail to determine the feasibility of this
option.

For the X-ray part of the mission, we explore the possibility to
include 8 X-ray telescopes in the spacecraft, taking advantage of the
technology developed for the eROSITA mission.  The result is an X-ray
survey with a flux limit of $8\times10^{-15}$ erg cm$^{-2}$ s$^{-1}$
(10 times more sensitive than DUO\cite{duo} and 5 times more sensitive
than eROSITA) that will identify all the clusters with a mass
$\gtrsim$ $2\times 10^{14}$ $\rm M_\odot$ up to $z \sim
1.5$. Therefore, the DEMON $5000 \rm deg^{2}$ X-ray survey is expected
to produce a catalog of $\sim 100000$ clusters with a mean redshift of
0.75.


The final spacecraft is 7.1 m high and 3.75 m wide (at its broadest
parts). It has an estimated mass of 2500 kg, which is well within the
transport capabilities of a Soyuz rocket into the desired LEO
orbit. The estimated cost of the spacecraft is 510 M euro~(see
Table~\ref{table:costs} for details).

The DEMON concept design shows that combining an optical and X-ray
survey in the same spacecraft is technically possible and very
interesting in terms of costs. In addition, it is able to provide high
quality data for both weak lensing and cluster statistics, something
that is not possible with the weak lensing missions proposed by
SNAP\cite{snap} and JEDI\cite{jedi} due to their poor sampling of the
FWHM PSF.

We also note that, although DEMON is optimized for a weak lensing
survey, the high quality of the observations and the wide sky coverage
of the survey will provide valuable data for other studies, like
strong lensing systems, baryonic wiggles, galactic structure or
detailed studies of clusters of galaxies. Due to the characteristics
of the X-ray instruments, DEMON will also contribute to the study of the
intracluster medium and galaxies with active nuclei
(AGNs). Furthermore, a number of faint objects will be identified as
future targets for narrow field instruments performing pointed imaging
observations, like XEUS.

Finally, the planned far-infrared surveys based on the
Sunyaev-Zel'dovich (SZ) effect (e.g. the ESA Planck mission and South
Pole Telescope) will produce the redshift dependent cluster mass
function and the power spectrum with different systematics compared to
the DEMON X-ray survey, permitting for a very important cross-check of
the two cosmological techniques to be made.

\acknowledgements       
 
The authors would like to thank the organizers and the tutors of the
Alpbach summer school 2005 for inspiration and help. Especially they
would like to thank Dr. Andreas Quirrenbach and the director of the
school, Dr. Johannes Ortner. We would also like to thank Edo Loenen
for help during the preparation of the 3D figure of the spacecraft,
and the Space Research Institute of the Austrian Academy of Sciences
and Dr. W. Baumjohann for the workshop support.  A.B.A. would like to
thank Dr. Jan Willem Pel and Dr. Isabel P\'erez Mart\'in for useful
discussions. F.M. would like to thank CEA-DAPNIA/SAp for technical
support and useful discussions during the preparation of the paper.

Financial support during the summer school, workshop and SPIE
conference was obtained from ANGLES, ASI, CEA-DAPNIA/SPP, INFN of
Bologna and Perugia, Danish Research Agency, DLR, Enterprise Ireland,
Institute of Astrophysics University of Innsbruck, Austrian Research
Promotion Agency FFG/ALR, GRICES, INAF, CIFS, NUI Maynooth
Postgraduate Travel Fund, SNSB, SRON and \"{O}sterreichische
Forschungsf\"{o}rderungsgesell\-schaft mbH.  The work by A.B.A. and
M.Z. was financed by the European Community's Sixth Framework Marie
Curie Research Training Network Programme under Contract
No. MRTN-CT-2004-505183 ANGLES.




\end{document}